\documentclass{sig-alternate-10pt}
\usepackage[margin=1in]{geometry}

\usepackage{graphicx,epsfig}
\usepackage{times}
\usepackage{dcolumn,bm,fleqn,epic,eepic,float}
\usepackage{amssymb,amsmath,multirow,rotate,color}
\usepackage{subfigure}
\usepackage[dvipsnames]{xcolor}
\usepackage[normalem]{ulem}
\usepackage[hyphens]{url}
\usepackage{paralist}
\usepackage{xspace}

\usepackage[T1]{fontenc}
\usepackage{textcomp}

\usepackage[smallbib,fnord,obey]{fixacm}

\newcommand{\ie}{\emph{i.e.} }
\newcommand{\aprx}{\raise.17ex\hbox{$\scriptstyle\sim$}}

\usepackage{color,soul}
\soulregister\cite7
\soulregister\citep7
\soulregister\ref7
\soulregister\href7
\soulregister\url7
\soulregister\texttt7
\soulregister\sout7
\soulregister\unichar7
\soulregister\vspace7
\soulregister\xspace7

\newcommand{\sw}{\textsc{SW}\xspace}
\newcommand{\arxiv}{\texttt{arXiv}\xspace}

\newcommand{\up}{\texttt{up}\xspace}
\newcommand{\bp}{\texttt{bp}\xspace}
\newcommand{\hirecs}{\texttt{HiReCS}\xspace}

\begin{document}
\pagestyle{empty}
%

\title{ScienceWISE: Topic Modeling\\ over Scientific Literature Networks}
%
%
%
%
%

\numberofauthors{14} 

\author{
%
%
A. Magalich, V. Gemmetto, D. Garlaschelli, A. Boyarsky\\
       \affaddr{University of Leiden, The Netherlands}\\
       \email{\{magalich, palchykov, gemmetto, garlaschelli, boyarsky\}@lorentz.leidenuniv.nl}
\and
A. Martini, A. Cardillo, A. Constantin, O. Ruchayskiy, P. De Los Rios, K. Aberer\\
       \affaddr{{\'E}cole Polytechnique F\'ed\'erale de Lausanne, Switzerland}\\
       \email{\{andrea.martini, alessio.cardillo, alex.constantin, oleg.ruchayskiy, paolo.delosrios, karl.aberer\}@epfl.ch}\\
\and
A. Lutov, M. Khayati, P. Cudr\'e-Mauroux\\
       \affaddr{University of Fribourg, Switzerland}\\
       \email{\{artem, mkhayati, phil\}@exascale.info}
\and
V. Palchykov\\
	\affaddr{University of Leiden, The Netherlands and Institute for Condensed Matter Physics, Liviv, Ukraine}\\
	\email{palchykov@lorentz.leidenuniv.nl}
}

\maketitle

\sloppy
\begin{abstract}
We provide an up-to-date view on the knowledge management system ScienceWISE (\sw) and address issues related to the automatic assignment of articles to research topics. So far, \sw has been proven to be an effective platform for managing large volumes of technical articles by means of ontological \textit{concept}-based browsing. However, as the publication of research articles accelerates, the expressivity and the richness of the \sw ontology turns into a double-edged sword: a more fine-grained characterization of articles is possible, but at the cost of introducing more spurious relations among them. In this context, the challenge of continuously recommending relevant articles to users lies in tackling a network partitioning problem, where nodes represent articles and co-occurring concepts create edges between them. In this paper, we discuss the three research directions we have taken for solving this issue:
\begin{inparaenum}[i)]
    \item the identification of generic concepts to reinforce inter-article similarities;
    \item the adoption of a bipartite network representation to improve scalability;
    \item the design of a clustering algorithm to identify concepts for cross-disciplinary articles and obtain fine-grained topics for all articles.
\end{inparaenum}
\end{abstract}

\section{Introduction}
\label{sec:intro}

The past several decades have seen a sizable growth of the global research community, as well as an acceleration of the scientific publishing workflow. This has led in turn to a drastic increase of the number of published works \cite{constantin2014automatic,nat-blog-2014}, a proliferation that has carried with it many technological challenges. One such challenge is the identification and discrimination of intra- vs. cross-disciplinary scientific production, in order to ease and enhance bibliographic search. The ability to make such a distinction is seen as dependent on the so-called ``conceptualizing of content'' \cite{Shibata2008, Waltman2012}, achievable through the maintenance of ontologies of concepts referred to in specialized literature. The huge amount of available articles, however, requires automating the construction of such ontological substrates as much as possible, but without compromising quality with respect to human-expert relevance judgments.

The ScienceWISE platform\footnote{\url{http://sciencewise.info/}} (henceforth indicated as \sw) aims at facilitating the necessary type of human-computer interaction to reach the above objective. It provides an interactive environment in support of the scientific community, where expert users can continuously contribute to the maintenance of a data-driven domain ontology through routine literature review activities. In doing this, they also give the system implicit feedback on content they consider relevant, which is subsequently used to provide better recommendations of new articles to bookmark and review. 

In this paper, we first provide an updated view of the \sw~platform and its mechanism for extracting concepts from scientific articles. We then look at aspects related to establishing a coherent correspondence between articles and broad topics of research. In particular, we will describe how the accurate selection of relevant concepts (Sec.~\ref{sec:rel_conc}), the choice of suitable representation schemes (Sec.~\ref{ssec:representations}) and accounting for overlapping community structure (Sec.~\ref{ssec:overlaps}) are all viable solutions to tackle the topic modeling issue. Finally, in Sec.~\ref{sec:conclusions}, we report on lessons learned and sketch future development possibilities.

\section{Platform Overview}
\label{sec:platform}

\begin{figure}[h!]
    \includegraphics[width=\columnwidth]{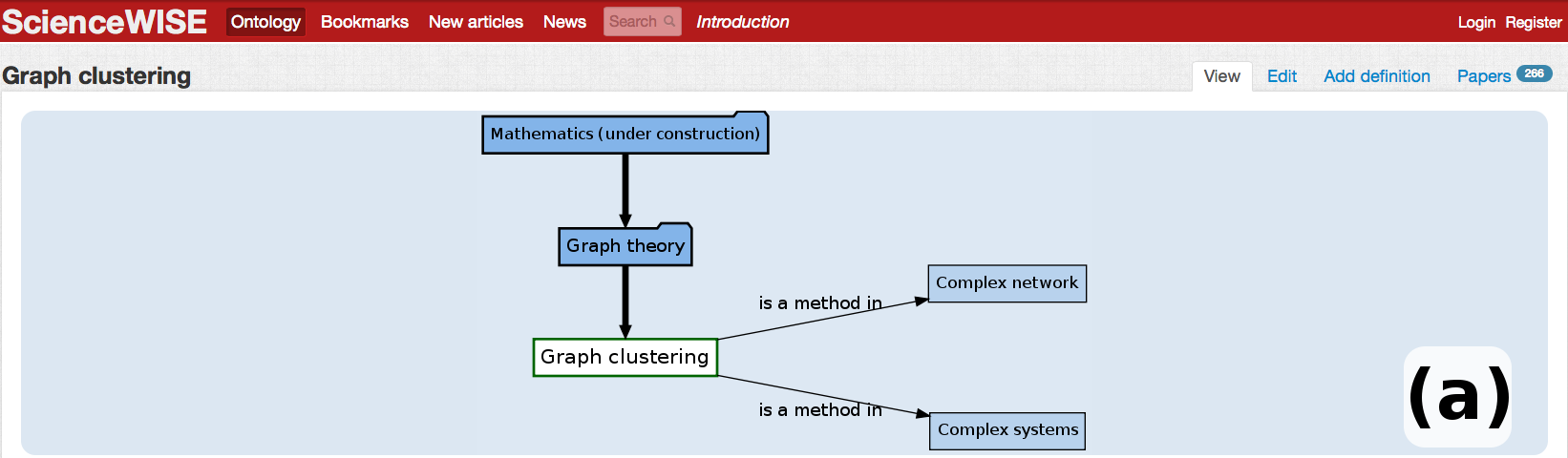}
    \caption{Ontological representation of the concept ``Dark Matter''. The entry includes information about categorization, related concepts, definitions and other resources.}		
    \label{fig:ontology}
\end{figure}

Organizing scientific literature in a systematic way calls for effective and efficient methods to constructing human-browsable taxonomies or ontologies that are field-specific, kept up-to-date and  interlinked to the contents of the documents they mean to represent. The extraction of relevant concepts from new publications thus constitutes the cornerstone of \sw. Currently, the platform's primary source of literature is the \texttt{arXiv.org} e-print repository of physical sciences articles \cite{ginsparg-nature-2011}, with which it synchronizes daily. Any newly published article is automatically analyzed for its most relevant concepts as well as novel terminology \cite{prokofyev2012tag,constantin2014automatic}. The article is then made available to the system's users for inspection and concept tagging. Through this process of bookmarking, users contribute expert input on the most relevant concepts of the article, possibly also specifying new concepts and semantic relations among them in the process, such as \textit{specialization} or \textit{similarity} (Fig.~\ref{fig:ontology}). 

\begin{figure}[h!]
    \includegraphics[width=1.02\columnwidth]{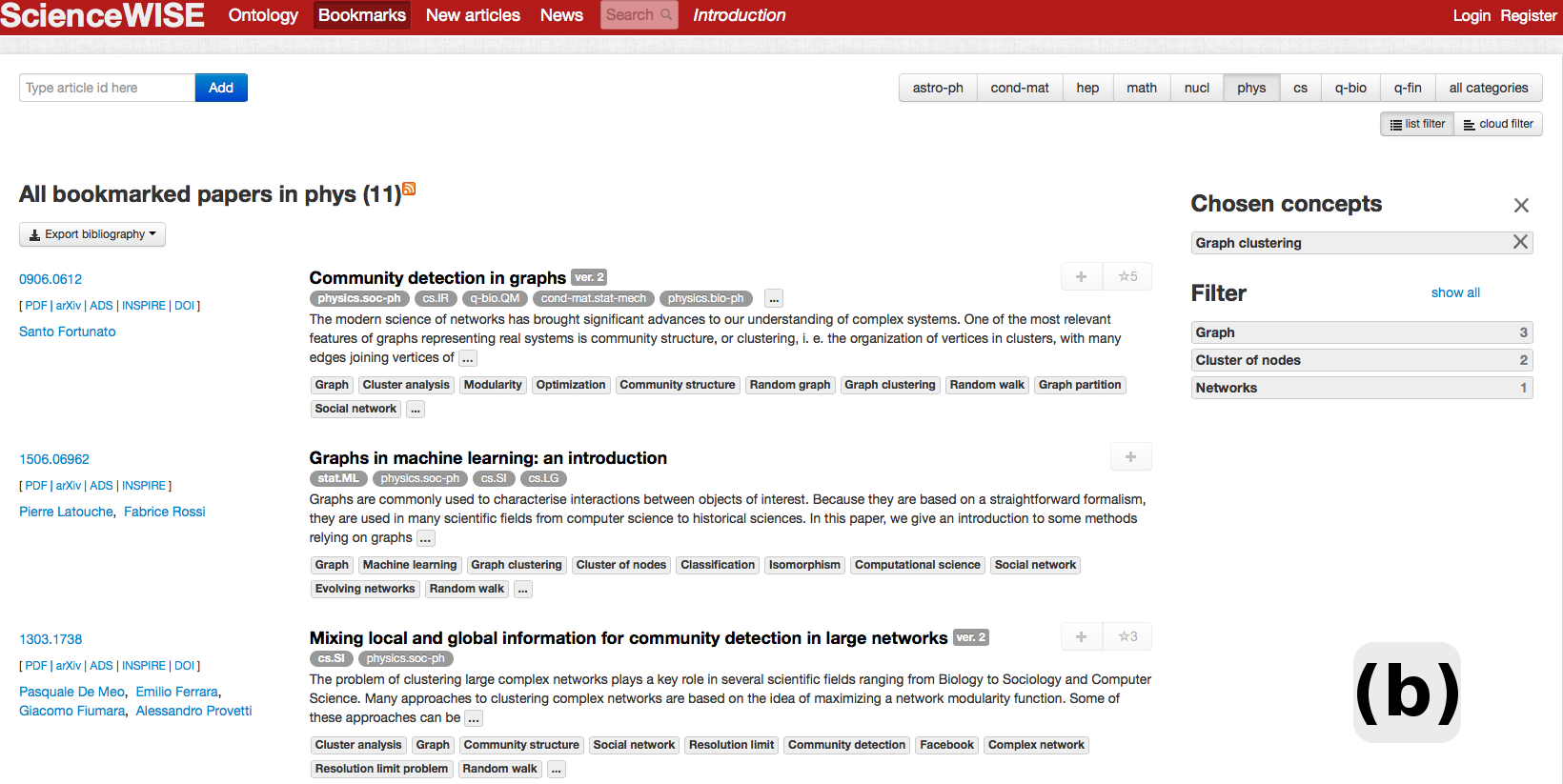}
    \caption{Article bookmarking interface. The column \textbf{Found concepts} contains ontological entries, \textbf{Possible concepts} -- novel extracted keyphrases, and \textbf{Chosen concepts} -- the user's personal choice.}
    \label{fig:bookmark}
\end{figure}

Upon bookmarking an article, the user is presented with three complementary options for selecting potential tags (Fig.~\ref{fig:bookmark}): 
\begin{itemize}
	\vspace{-3pt}
	\itemsep0em
	\item choose existing ontological concepts found in the article;
	\item choose automatically extracted keyphrases, that are not currently part of the ontology;
	\item manually add any missing but relevant terms.
\end{itemize}

The user may choose any combination of the above options to compile a set of terms that best represents his/her interest in the article in question. For convenience, the list of ontological concepts is ranked by an enhanced TF-IDF measure \cite{prokofyev2012tag}, while the list of keyphrases is generated and ranked using the KPEX algorithm \cite{constantin2014automatic}. New user contributions (concepts and/or relations) are periodically validated by a group of system curators (power-users) before being permanently adopted as updates to the current ontology.

As an additional feature, bookmark collections may be analyzed to provide each user with personalized article recommendations. Since scientific literature is generally quite specific, the co-occurrence of concepts within the same article often suggests a thematic relation between them, whilst the presence of the same concept in different articles is a building block to gauge the relation between different articles. This feature favors the use of \sw as a personalized semantic filter over the stream of new, incoming articles. It is at this point, however, that new methods for deriving \emph{meaningful relations} between articles are needed. 
The information extracted from \sw can, in turn, be represented as a set of elements belonging to two distinct classes: articles and concepts, which can be encoded as a \emph{complex network} \cite{boccaletti-phys_rep-2006, newman-book-2010}. 
A network representation can be very effective in highlighting important relations. When the underlying ontology becomes too rich, however, the probability that two articles share one concept by chance becomes non-negligible. This phenomenon is responsible for the emergence of many spurious interactions that raise the density of connections and consequently makes the information extraction process harder. We had to devise a number of new solutions to work around this network density problem in \sw. The following sections showcase three successful but complementary ways to address this issue, and that are applicable at different stages of the processing pipeline -- a better selection of relevant concepts, a more effective network representation and a more meaningful partitioning of the network.

\section{Entropic Identification of Relevant Concepts}
\label{sec:rel_conc}



Common or generic concepts present in most articles (such as \emph{model} or \emph{star} in physics) create many spurious interactions in a network representation. The resulting structure is thus very dense, resembling a complete graph where all nodes (articles) have a near-direct link to each other. 
Several methods have already been proposed to tackle the problem of dropping some of the links in dense networks to make the networks sparser \cite{serrano-pnas-2009,radicchi-pre-2011} but they operate in an \textit{ex-post} way, validating connections in a resulting network based on their weights. In our context, the challenge is instead to reduce the number of connections and the effect of spurious similarity on the weights \emph{a priori}, by filtering away the common (or generic) concepts, which do not provide any valuable classification information.  This is already achieved in \sw~by a crowdsourced concept classification, i.e. through the assignment of a \textit{generic concept} flag. Yet, the concepts that may be considered as generic in one discipline may be quite specific in another. This makes the problem of identifying common concepts strongly dependent on the considered collection of papers and their topics. Hence, the problem requires a solution that does not rely on human-made input, but rather supplements it.

\begin{figure}[ht!]
\centering
\includegraphics[width=\columnwidth]{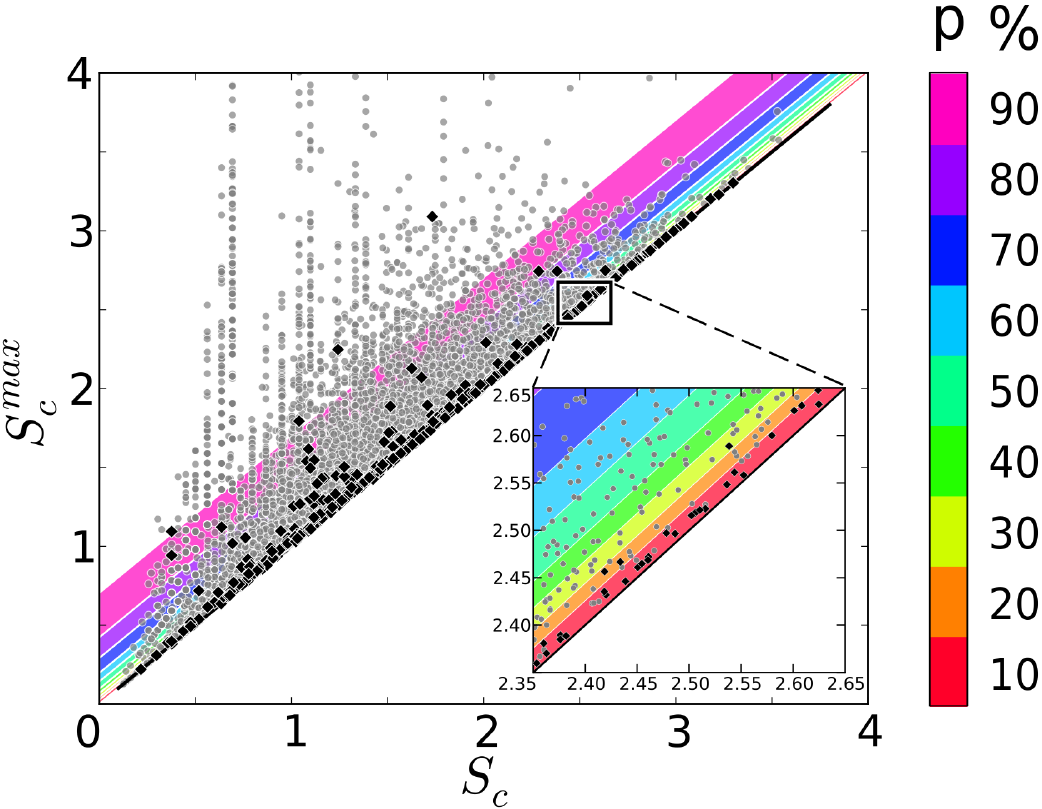}    
\caption{Relation between the entropy $S_c$ of a concept $c$ and $S_c^{\max}$, for the collection \texttt{arxivPhys2013pc}. The background colors delimit the regions where the distance between a point and the line $S_c=S_c^{\max}$ is less than a given percentile $p$. Black dots identify the concepts marked as \emph{generic} by experts.}
\label{fig:entropy}
\end{figure}

Our solution to this problem consists in making use of \emph{entropy} to select relevant concepts. Entropy has previously been used in semantic analysis studies to classify documents \cite{hotho-jcllt-2005,berger-complin-1996} based on the frequency of words in a text -- \ie \emph{term frequency} $TF$ --  \cite{zipf-book-1949,ferrer_i_cancho-pnas-2003,font_clos-njp-2013,gerlach-njp-2014}. Specifically, we make use of Shannon's entropy to quantify the amount of information carried out by concepts. For each concept $c$, we compute two different entropies: one, $S_c$, associated to the observed $TF$ probability distribution, and the other, $S_c^{\max}$, associated to a probability distribution derived from the \emph{maximum entropy principle} and constrained by the first moment and log-moment of the observed $TF$ distribution \cite{berger-complin-1996,baek-njp-2011,yan-pone-2015}. The maximum entropy value acts as an upper bound for the observed entropy, allowing to determine if the observed value is big or small with respect to a quantity computed for the same concept. Hence, the comparison of the two entropy values (Fig.~\ref{fig:entropy}) constitutes the core of the methodology we used to filter the similarity network.

More specifically, in Fig.~\ref{fig:entropy} we consider the distribution of points in the $S_c$ \textit{vs} $S_c^{\max}$ space for the concepts extracted from all the articles submitted during the year 2013 to the \arxiv in \texttt{physics} subject classes as principal category (hereafter \texttt{arxivPhys2013pc}), which comprises nearly 30k articles. A glance at the diagram reveals some intriguing features. For instance, the vast majority of hand-labeled generic concepts (black dots) sit along the $S_c=S_c^{\max}$ line, suggesting that generic concepts tend to maximize their entropy and, consequently, carry little information. However, a more careful analysis (displayed in the inset)  hints at the presence of many more concepts whose observed entropy is close to the maximum. This paves the way for the establishment of a parallel between generic concepts and those with an observed entropy closer than a given threshold $p$ to their maximal one $S_c^{max}$.

The second interesting feature is the presence of a few groups of common concepts located away from the boundary. These outliers have been marked by users who might correctly consider them as common in their own sub-discipline, but which might represent only a small fraction of the whole dataset. On the network density side, the pruning of basic concepts allows to drop significantly the link density. As an example, by removing concepts falling within the 10th percentile, we are able to decrease link density by \aprx20\%. Such a drastic reduction leads to a clearer and more specific community structure and topic classification. Furthermore, the automatic identification of common concepts can thus be used to signal potentially wrongly-defined generic concepts.

\section{Alternative Graph Representations}
\label{ssec:representations}

\begin{figure}[!ht]
\centering
\includegraphics[width=\columnwidth]{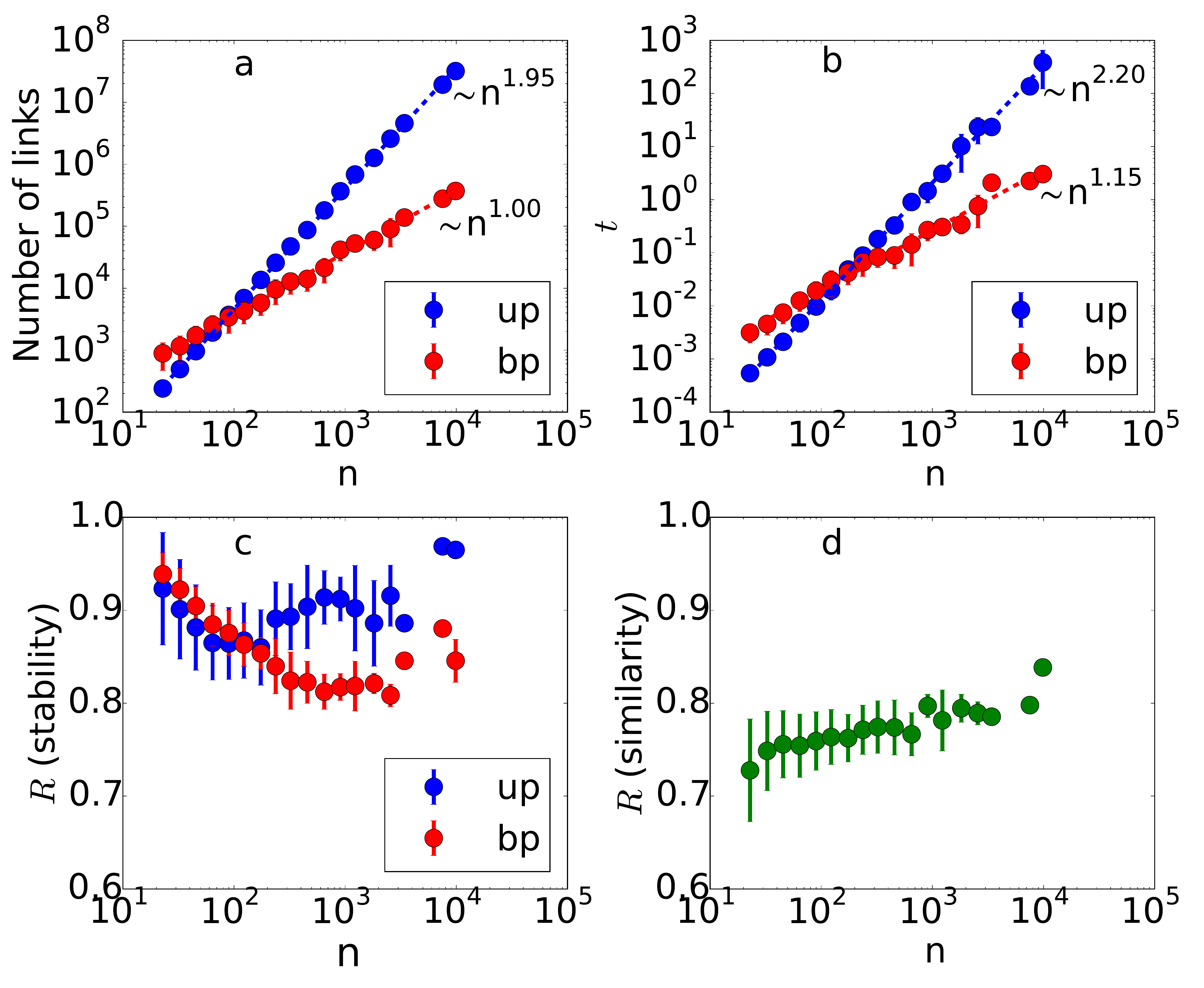}    
\caption{(a) Number of links as a function of the network size $n$. (b) Run-time required to compute one single realization of the Louvain algorithm as a function of $n$; results refer to its \texttt{python-igraph} implementation. (c) Stability of the outcome of the stochastic algorithm for the two different network representations. (d) Similarity between the \up and \bp partitions.}
\label{fig:Louvain}
\end{figure}

The set pf articles and concepts used by \sw has thus far been represented as a unipartite (\up) network, where only articles are nodes. 
Alternatively, the system may be represented as a bipartite (\bp) network, where its articles and concepts can be directly mapped to two types of nodes. Links (edges) only connect cross-type nodes -- articles to their respective concepts. 
Such an alternative representation eliminates the density problem since the number of links in a bipartite network grows linearly with the number of articles, $n$, while it grows as $O(n^2)$ in the unipartite case (see Fig.~\ref{fig:Louvain}(a)).
Consequently, the bipartite approach significantly reduces the computational resources needed to cluster big collections without loss of quality, as shown below.

Besides the \texttt{arxivPhys2013pc} dataset, we also consider 1084 of its subsets of different sizes accounting for single \arxiv categories ($7$k -- $11$k articles);  sub-categories ($280$ -- $3$k articles) and smaller groups of papers (with at least $20$ nodes) that are rather focused on specific topics in their own disciplines. All these subsets have been selected \textit{uniformly} by recursively applying modularity optimization methods~\cite{Blondel2008} to the \up representation of the entire collection and the resulting clusters/sub-sets~\cite{Palchykov2016}. 

We investigated run-time dependence on the size of the dataset for the \up and \bp representations using the Louvain clustering algorithm~\cite{Blondel2008}, which aims at maximizing the modularity function~\cite{newman2004}. Originally, the modularity function had been designed for unipartite networks. To deal with bipartite networks, Barber's generalization of modularity~\cite{Barber2007}, which adjusts the underlying null model, was employed. 
However, our preliminary examination shows that the choice of the null model is not crucial, as it barely affects the community detection results. This motivates us to use the same implementation in order to make a consistent comparison between the resources required to cluster articles in the two representations.

The run-time results are showcased in Fig.~\ref{fig:Louvain}(b), where the average time to cluster a collection of a given size is supplemented by the standard deviation. In both representations, the run-time $t$ scales as a power-law function of the size $n$ as $t \aprx n^\alpha$. However, the scaling is close to quadratic for the \up representation ($\alpha\approx2.20$), but about linear for the \bp one ($\alpha\approx1.15$). The difference is likely due to the aforementioned discrepancy in the number of links between the two representations, which follows roughly the same trend. For small datasets (below 100 articles), a unipartite approach is more convenient, while for bigger collections the bipartite representation prevails. Noticeably, there is a 100 times difference in the run-time for collections of $n=10^4$ articles.

Since the heuristic used be the Louvain algorithm has a stochastic component, the results of different runs may, in general, differ. Thus, it is crucial to understand how stable the outcomes of different runs are: is it necessary to consider an optimal partition over many runs or is it enough to take just one run? To answer this question, we first identify 100 partitions for each dataset and each network representation. Then, we consider all possible pairs of these outcomes and perform pairwise comparisons using Rand Index $R$ \cite{Rand1971}. The behaviors of $R$, shown in Fig.~\ref{fig:Louvain}(c), indicates that a single run of Louvain suffices for any \up network and for small \bp networks. 

Although stability is a fundamental requirement, it is also important to establish how similar two partitions are within different representations. We therefore compared each of the 100 partitions of \up representations with each partition obtained from the \bp representation for every single considered dataset. The results in terms of $R$ are shown in Fig.~\ref{fig:Louvain}(d): significant discrepancies for small datasets have been observed, while the two representations lead to similar outcomes for larger collections.
Comparing these results to the stability results (Fig.~\ref{fig:Louvain}(c)), we conclude that the similarity between \up and \bp partitions is closely approaching the (smallest) stability index of the resulting partitions for all datasets above 100 articles. This means that the clustering obtained using the two alternative approaches are similar, though this similarity is limited by the stability restrictions of each approach.

To summarize, our resource and quality performance tests have highlighted under which conditions \bp and \up approaches may be used as alternative ways to cluster article collections with comparable outcomes. There is a clear trade-off between computational resources and stability: smaller datasets should be clustered using the \up representation (good stability, low resource consumption) while larger datasets should employ \bp clustering despite less stable results.

\section{Overlapping communities}
\label{ssec:overlaps}

In the above sections, we considered the non-overlapping clustering of articles where each manuscript belongs to a single topic. However, there are many cross-disciplinary articles whose contents do not belong to a single well-defined topic. In this section, we therefore consider overlapping clustering~\cite{Lcn11}, and instead of maximizing the usual modularity function, we maximize its extended version \cite{nicosia2009}.

In this context, we propose a greedy optimization algorithm, \hirecs, inspired by the Louvain routine. At each single iteration it merges initially ungrouped nodes into communities using maximal a Mutual Modularity Gain (MMG) criterion. 

More specifically, for each node \verb|#i| in the network, we identify a set of neighbors such that the pair-wise merge of \verb|#i| with each of these leads to a maximal gain in modularity. Then, two nodes \verb|#i| and \verb|#j| are merged into a cluster if such a move is optimal for both of them.

An overlap appears when for a given node \verb|#i| several neighbors satisfy the MMG criterion equally. To represent the overlap in the shared node \verb|#i|, we replace \verb|#i| by a set of replica nodes~\verb|#ik|, one for each of the $K$ overlapping clusters $C_k$. To ensure the conservation of the weights of the links incident to a shared node, we equally distribute them across all the replicas as follows:
\begin{equation*}
\forall\: k \in \{1,\ldots,K\};\quad w_{ik,j} =  \frac{w_{i,j}}{K} \,,
\label{eq:ovpres_wji}
\end{equation*}
where $w_{ik,j}$ stands for the weight of the link between \verb|#ik| and \verb|#j|. Unlike the Louvain algorithm, \hirecs groups the nodes into fine-grained clusters in a deterministic way.

The clustering itself is an agglomerative hierarchical method, 
where each clustering iteration builds a subsequent level of the hierarchical community structure. At each iteration, the non-clustered nodes are propagated to the next iteration where they are treated as input nodes, as well as the previously formed clusters.
The execution terminates when a new iteration does not produce any new cluster.


To assess the quality of the results, our algorithm was applied to synthetic networks with overlapping clusters generated by the \textit{LFR} benchmark~\cite{Lcn09b}. We compare \hirecs against state-of-the-art algorithms, including Louvain~\cite{Blondel2008}, Ganxis~\cite{Xie11}, Oslom2~\cite{Lcn11},  SCP~\cite{Kpl08} and a random communities generator (Rcoms). The latter takes the number and size of clusters from the ground-truth and assigns connected nodes randomly to clusters.
\begin{figure}[ht]
  \centering
  \includegraphics[scale = 0.5]{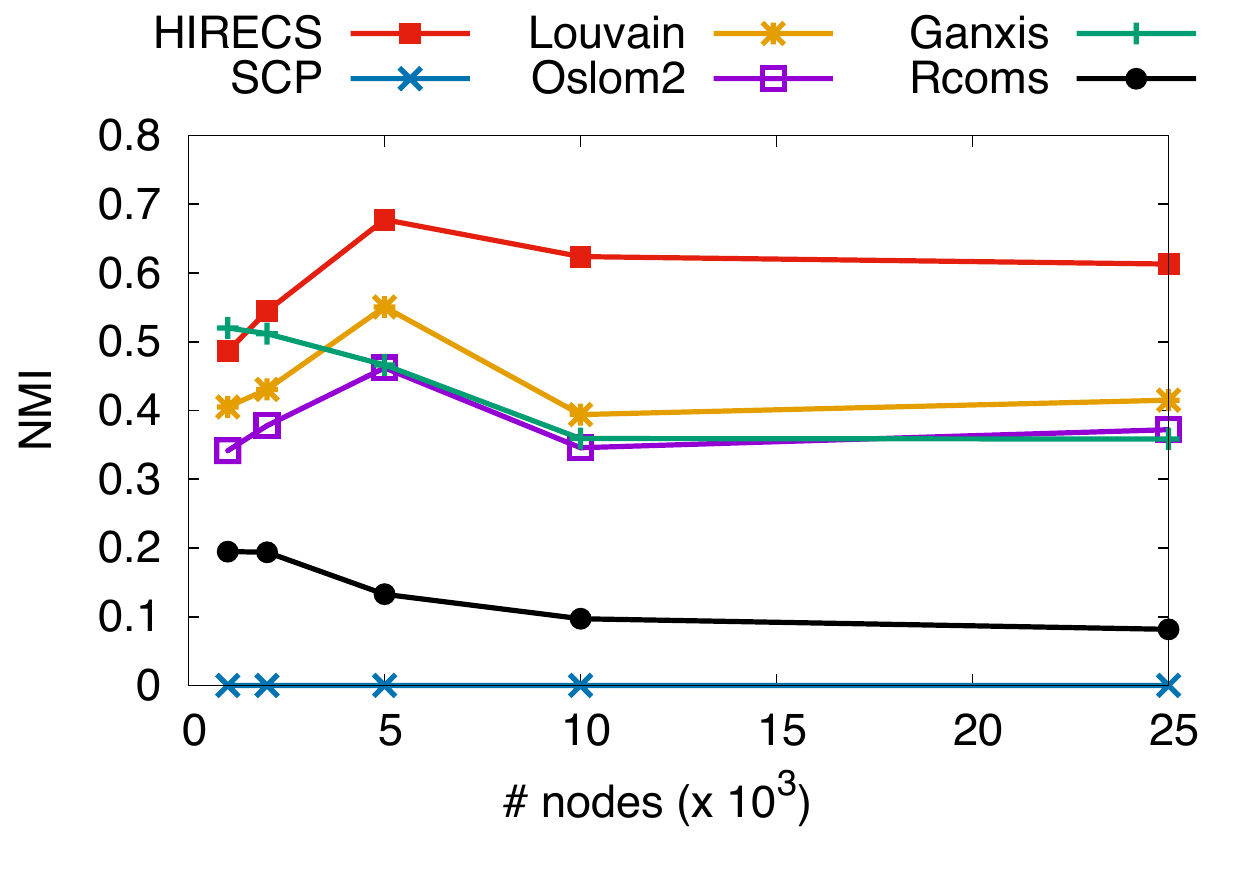}
  \caption{Comparison of NMI values achieved by different clustering algorithms, as a function of the number of nodes in the benchmark network. Results are averaged over eight network instances.}  
  \label{fig:nmi}
\end{figure} 

To assess the similarity between the resulting clusters and the ground-truth communities, we use the Normalized Mutual Information (NMI) for overlapping clusters~\cite{Esv12}, which is fully compatible with the standard NMI~\cite{manning2008introduction}. The NMI values vary in the range $[0,1]$, where a value of 1 signifies that a clustering is identical to the ground-truth.
The results shown in Fig.~\ref{fig:nmi} show that \hirecs outperforms state-of-the-art solutions, including both parameter-free and parameterized ones on sparse networks, due to the fine-grained structure of the clusters resulting from the clustering.

Fine-grained cluster identification is important to trace and interrelate topics from the most common ones to the most specific ones, as well as to identify topics in cross-disciplinary articles. 

Those experimental results show that our algorithm grasps the community structure better than its competitors in a parameter-free way and allows to identify multiple topics for a given paper. 
Additionally, \hirecs achieves a linear space complexity and near-linear run-time complexity on sparse networks with respect to the number of links.

\section{Conclusions}
\label{sec:conclusions}

\subsection{Summary}
In this paper, we showcased the novel methods that \sw employs to address the problem of semantic organization and contents retrieval on large collection of scientific papers. Since its launch in 2009, the system has reached several milestones, evolving from the whim of a few scientists to a well-established platform used by hundreds of users on a daily basis. Based on the experience accumulated by developping and using \sw~over the years, we are able to pinpoint some important lessons learned so far, and to use them as departure points for the further developments of the system. In this article, we highlighted three key issues related to the assignment of topics to document networks. More specifically, we showed how:
\begin{itemize}
 \item The automatic selection of relevant concepts based on entropy favors the emergence of strongly-organized structures -- also at the level of sub-topics -- and, more importantly, does not rely on human evaluation/validation.
 \item The bipartite representation of the system can be used to overcome the limitations imposed by the density of the unipartite graph. Moreover, the stability of the results, albeit being inferior to the unipartite case, remains still high.
 \item The \hirecs algorithm, which supports overlapping communities, 
 can be applied to improve both the classification of cross-disciplinary articles and the identification of fine-grained topics for the articles.
 %
\end{itemize}

\subsection{Future Work}
\label{ssec:future}

A number of important challenges lay ahead. We are currently working on overtaking the current limitations of the methodologies presented in this manuscript, for example, by combining together two -- or more -- approaches. Another challenge relates to the application of the insight extracted from physics-related articles (for which a well-established ontology exists) to other disciplines such as biomedicine or environmental sciences, and to non-scientific media altogether, where ontologies are scarce or even often missing. In a wider perspective, \sw can be seen as an ideal ecosystem to test various search, clustering and data management technologies, ranging from community detection and topic modeling, entity linking or knowledge extraction.

\section*{Acknowledgments}
This work was supported by SNSF project No. 147609 (Crowdsourced conceptualisation of complex scientific knowledge and discovery of discoveries) and by the EU project MULTIPLEX (contract 317532).

\bibliographystyle{abbrv}
\bibliography{sigproc}  

\end{document}